# A hybrid origin for the Martian atmosphere

Kaveh Pahlevan[1*], Laura Schaefer[2], Don Porcelli[3]

1. Carl Sagan Center, SETI Institute, Mountain View, CA, USA

2. Department of Geological Sciences, Stanford University, Stanford, CA, USA

3. Department of Earth Sciences, Oxford University, Oxford, UK

*To whom correspondence should be addressed:

Email: kpahlevan@seti.org

Tel: +1 (480) 401 8584

3273 words (abstract, acknowledgements & captions included)

30 references

4 Figures

Revised for publication in Geochemical Perspectives Letters

Keywords: Mars, primordial outgassing, atmosphere, nebular capture, hydrodynamic escape

**Abstract**

The Martian isotopic record displays a dichotomy in volatile compositions. Interior volatiles from the mantle record a chondritic heritage (*e.g.*, H, N, Kr, Xe) whereas the atmospheric reservoir of Kr and Xe – which do not currently experience escape – record heritage from a solar-like source. Motivated by disparate inferences on the source of Martian atmospheric volatiles (outgassed versus nebular captured), we consider hybrid-source accretionary atmospheres in which a high molecular weight (*e.g.*, $CO_2$-rich) outgassed component is mixed in with the low molecular weight $H_2$-rich nebular atmosphere. We conduct calculations of nebular capture with and without a high molecular weight outgassed component mixed into the atmosphere during the lifetime of the solar nebula. Mixing an outgassed component into the nebular layer enhances the captured gas inventory by ≈1-3 orders of magnitude – depending on the outgassed inventory – relative to "pure" nebular capture. These observations and calculations suggest that the Martian atmosphere arose as a subequal mixture of outgassed and nebular-derived components, and provide a framework for assessing the role of various mechanisms of gas loss over the entire history of the planet.

**1. Introduction**

Mars has recently emerged as a natural laboratory for studying the acquisition and processing of volatile elements on the terrestrial planets. Because undifferentiated building blocks of terrestrial planets were likely volatile-bearing (SCHAEFER AND FEGLEY, 2017) and Martian accretion likely occurred in the presence of the volatile-rich solar nebula (DAUPHAS AND POURMAND, 2011), questions about different volatile sources and formation processes for the primordial Martian atmosphere can now be addressed (SAITO AND KURAMOTO, 2018; PÉRON AND MUKHOPADHYAY, 2022). High-precision isotopic measurements display a dichotomy in the sources of Martian volatiles. Volatiles dissolved in melts derived from the Martian mantle are observed to have an isotopic composition akin to chondrites for hydrogen, nitrogen, krypton and xenon (USUI, 2019; PÉRON AND MUKHOPADHYAY, 2022; DELIGNY et al., 2023) whereas the isotopic composition of krypton and xenon in the Martian atmosphere – which do not currently experience escape – record a solar-like source composition (PEPIN, 1991; CONRAD et al., 2016). These observations prompt questions about the relationship between Martian interior and atmospheric volatiles.

The volatile dichotomy between a chondrite-like mantle and a solar-like atmosphere has recently been interpreted to mean that the Martian atmosphere cannot be the result of magma ocean outgassing and must be the result of gravitational nebular capture (PÉRON AND MUKHOPADHYAY, 2022). There is, however, empirical isotopic evidence that silicate Mars experienced outgassing during the lifetime of the short-lived and volatile radionuclide $^{129}$I ($\tau_{1/2}$=16 Myrs), resulting in the observed $^{129}$Xe-depletion in Martian interior reservoirs (MARTY AND MARTI, 2002). Early selective removal of volatile elements from silicate Mars – for example via outgassing from a molten state (ELKINS-TANTON, 2008) – points towards the transport of chondritic volatiles into the primordial atmosphere. Although this episode of primordial outgassing is empirically supported, it has yet to be reconciled with the observed dichotomy between the Martian interior and atmospheric volatiles (PÉRON AND MUKHOPADHYAY, 2022). Here, we consider hybrid accretionary atmospheres (SAITO AND KURAMOTO, 2018) in which an outgassed high-molecular weight gas is mixed into the distended low-molecular weight nebular atmosphere. We show that such mixing can reconcile primordial outgassing of chondritic volatiles with nebular capture of solar-like gases into the Martian atmosphere, that mixing before dissipation of the solar nebula strongly enhances the mass of nebular captured gas, and that comparison of hybrid-source compositions with that of the present-day atmosphere can yield new insights into the history of Martian atmospheric evolution. As an example, the hybrid origin model implies that most Martian argon loss occurred primordially rather than during later geologic epochs as commonly assumed. We show that a hybrid initial composition followed by an early episode of hydrodynamic escape is consistent with the observed heavy noble gas abundances (Ar/Kr) and is a viable initial condition for understanding Martian atmospheric history.

## 2. Model and Results

***Enhancement of nebular capture via mixing.*** We calculate structures for dusty (Eq. S-4) captured Martian atmospheres present during accretion in hydrostatic equilibrium and thermal steady state with the solar nebula. We consider atmospheres both with and without mixing of an outgassed high mean molecular weight layer (see Supplementary Information for discussion on the energetics of mixing via thermal convection). Atmospheres without a high molecular weight component are called "pure" and are presented as a reference. The atmospheres are convective at depth and radiative at altitude, and are assumed to blend into the solar nebula at the Hill radius at ≈320 Mars radii. Because the mass (and heat capacity) of these atmospheres is relatively small, a heat source other than secular cooling is needed to calculate quasi-static structures. An absolute lower limit on the heat flow derives from long-lived radioactive decay and is equivalent to accretion rates ≈$10^{-4}$ Mars masses/Myr (ERKAEV *et al.*, 2014). More likely heat flows relevant to the first few million years of Martian history derive from ongoing planetesimal accretion, which also delivers high molecular weight volatiles, and/or $^{26}$Al decay. We consider heat flows at the base of the atmosphere equivalent to planetesimal accretion rates of 0.01-1 Mars masses/Myr (Eq S-1), covering the range from energetic accretion consistent with large-scale melting (DAUPHAS AND

POURMAND, 2011) down to reduced heat flows unable to maintain a magma ocean (SAITO AND KURAMOTO, 2018) and more consistent with sweep-up of planetesimals during the waning stages of accretion. For heat flows equivalent to accretion rates of 0.01-1 Mars masses/Myr, the mass ($M_H$) of a "pure" nebular captured atmosphere is ≈0.2-3.1x10$^{18}$ kg and equivalent to ≈0.05-0.8 bars of $H_2$ at the Mars surface (Fig. 1). Lower accretion rates produce cooler, denser atmospheres that are also more massive, a behaviour summarized with the maxim: "to cool is to accrete" (LEE AND CHIANG, 2015).

Mixing of the nebular atmosphere with an outgassed high molecular weight gas – for which there is sufficient energy (Eq. S-13) – sharply enhances the mass of the captured gas inventory. To illustrate the magnitude of this effect, we calculate the structure of hybrid-source atmospheres in which the outgassed component consists of pure $CO_2$ ($\mu$=44 amu with an inventory size characterized by $M_{CO2}$) with the nebular component ($\mu$=2.4 amu with an inventory size characterized by $M_H$) assumed to be fully-mixed into the outgassed layer, producing a homogeneous hybrid-source atmosphere (see equations S-1 to S-3 and S-7 to S-12 in the Supplementary Information). To find solutions, we take the mean molecular weight of the mixture as a free parameter, varying it across a range ($\mu$=6.6-19 amu) that allows sampling of the solution space. When a well-mixed hybrid atmosphere hydrostatically equilibrates with the solar nebula and achieves thermal steady state, the mass of the captured inventory ($M_H$) strongly depends on the inventory of the high molecular weight gas ($M_{CO2}$) with which it is mixed (see Gas-assisted capture, Fig. 1). For planetesimal accretion rates of 0.01-1 Mars masses/Myr and heavy gas inventories ($M_{CO2}$=3.9-390x10$^{19}$ kg) equivalent to 10-1000 bars of $CO_2$ at the planetary surface (ELKINS-TANTON, 2008), the captured nebular inventory ($M_H$) is in the range of 1.2-120x10$^{19}$ kg, equivalent to ≈3-300 bars of $H_2$ at the surface (Fig. 1). The strong enhancement of nebular gravitational capture via mixing an outgassed component is mainly due to an increase in mean molecular weight, which increases the gravitational coupling between the atmosphere and planet and decreases the atmospheric scale height, which – like cooling – causes contraction and an increase in the hydrogen density of the lower atmosphere where most of the atmospheric mass resides. This behaviour can be summarized with another maxim: "to mix is to accrete."

The dependence of the nebular captured inventory ($M_H$) on the outgassed inventory ($M_{CO2}$) in fully-mixed hybrid-source atmospheres (Fig. 1, and Eq S-11 and S-12) sets upper limits on the contribution of the nebular component to the Martian atmosphere. A measure of the relative contribution of captured and outgassed inventories is the mean molecular weight of the mixture, which only varies by a factor of ~three for the full range of conditions that we consider (Fig. 1). Although the mass of the outgassed component is dominated by carbon species (*e.g.*, $CO_2$), other outgassed volatiles (H, N, and noble gases) were also present and the elemental abundances and isotopic composition of the hybrid mixture can be used – in comparison with observed abundances – to further constrain the nebular contribution to Martian volatiles. Whereas the physics of nebular capture into fully-mixed hybrid atmospheres yields upper limits on the nebular contribution, lower limits can be derived from the cosmochemistry of chondritic-nebular gas mixtures, as we show in the following section.

***Chondritic-nebular gas mixtures.*** To describe cosmochemical consequences of mixing a nebular component into an outgassed atmosphere, we calculate two-component mixtures including major volatiles (H, C, N) and noble gases (Ne, Ar, Kr, Xe) with chondritic and nebular endmembers. We neglect He because – like He in Earth's atmosphere – the lifetime of this noble gas with respect to escape from the Martian atmosphere is thought to be extremely short relative to geological timescales and its present-day abundancemay simply reflect a balance between recent supply and loss (KRASNOPOLSKY *et al.*, 1994). For the chondritic endmember, we adopt the 55% H chondrite 45% EH chondrite model for Mars (SANLOUP *et al.*, 1999). For the purposes of the mixing calculations, we include all chondritic volatiles (interior and outgassed), although a substantial fraction of chondritic H (or "water") is expected to remain sequestered in the interior (SIM *et al.*, 2024). The assumption of complete outgassing may be more accurate in the case of C, N, and the noble gases. With endmember compositions specified, the composition of the resulting mixture can be described with one parameter, the relative contribution of the two components.

The compositional characteristics of a nebular-chondritic mixture can be calculated for comparison with the observed atmospheric abundances. For "Gas assisted capture" (Fig. 1), representing fully-mixed hybrid-sourced atmospheres in hydrostatic equilibrium with the solar nebula, the nebular contribution to the total volatile budget of Mars, counting atoms, is ≈46-77%, depending on the accretion rate and outgassed volatile inventory. The total volatile budget is dominated by H and He from the nebular component and H and C from the chondritic component. For these relative proportions of the nebular component to the hybrid mixture, the neon (>99%), argon (>99%) and krypton (>90%) inventories are dominated by the nebular component, whereas carbon (<1%) and nitrogen (<2%) inventories experience negligible nebular additions and continue to be dominated by the outgassed component (Fig. 2). Hydrogen and xenon are intermediate cases in which the inventories in the resulting mixture are derived from comparable contributions from the two sources. In summary, nebular capture via complete mixing into a hybrid-source atmosphere produces Martian Ne, Ar and Kr with solar heritage, C and N with chondritic heritage, and H and Xe with mixed heritage.

A hybrid primordial mixture is consistent with the observed isotopic composition of the Martian atmosphere. Mass-selective loss has fractionated stable Ar isotopes ($^{36}$Ar from $^{38}$Ar) but a solar-like source is a viable starting composition for atmospheric argon (ATREYA *et al.*, 2013). Krypton in the atmosphere is isotopically distinct from chondrites but nearly indistinguishable from solar(PEPIN, 1991; CONRAD *et al.*, 2016). Atmospheric xenon can be modelled either as mass-fractionated solar or mass-fractionated chondritic gas (SWINDLE, 2002). Although the physics of nebular capture into fully-mixed hybrid atmospheres sets upper limits on the nebular contribution (<46-77%), partial mixing could yield a lower nebular contribution. The requirement that Martian Kr be nearly indistinguishable from solar but clearly distinct from chondrites conservatively constrains the nebular contribution to the total Martian volatile budget to >10%, counting atoms (Fig. 2). Next, we consider the consequences of primordial hybrid mixtures for inferring Martian atmospheric history.

*Elemental abundances elucidate escape processes.* Of all the major volatiles (H, C, N) and noble gases (Ne, Ar, Kr, Xe) we consider, krypton is most nearly isotopically unfractionated in Mars's atmosphere relative to its apparent source, the solar nebula. Accordingly, to gain insight into the nature of evolutionary processes, we consider elemental abundances normalized to krypton and relative to solar elemental abundances. Hybrid-source elemental abundances have some affinity to modern Mars (Fig. 3), with important differences. Relative to hybrid-source mixtures, the modern Martian atmosphere is elementally depleted in H, C, N, Ne, Ar, and Xe, each of which is also enriched in the heavy isotopes in the Martian atmosphere (BOGARD et al., 2001). Such coupled elemental and isotopic fractionation suggests the viability of a hybrid-source mixture as a precursor to the modern Martian atmosphere, the two being linked via compositional evolutionary processes, among which mass-selective losses to space looms large.

We consider elemental ($^{36}$Ar/$^{84}$Kr) and stable isotopic ($^{36}$Ar/$^{38}$Ar) fractionation accompanying argon loss. Argon is suitable for examining ancient processes because the inventory of atmospheric $^{36}$Ar is primordial meaning it cannot be accounted for by volcanic outgassing over time (JAKOSKY AND TREIMAN, 2023). The non-radiogenic Ar/Kr ratio in the modern Martian atmosphere is lower than that of a hybrid mixture by a factor of ≈50 (Fig. 3) whereas the $^{36}$Ar/$^{38}$Ar is only lower than plausible sources by ≈25% (ATREYA et al., 2013). A loss process that strongly separates Ar from Kr but only weakly discriminates $^{36}$Ar from $^{38}$Ar is indicated. We consider an episode of extreme ultraviolet (EUV) powered hydrodynamic escape in which an outflow of $H_2$ and $CO_2$ entrains trace gases via frequent collisions (ZAHNLE et al., 1990). This loss process is thought to be the main way by which $H_2$-dominated atmospheres dissipate.. Entrainment involves all trace gases up to a maximum molecular mass whose value depends on the strength of the escape flow (See Eq S-16 in the Supplementary Information for details). Model results reveal the existence of a hydrodynamic outflow sufficiently strong to reproduce the chemical (Ar/Kr) and isotopic ($^{36}$Ar/$^{38}$Ar) fractionation observed in the Martian atmosphere starting from a hybrid mixture while remaining sufficiently weak to allow Kr to be retained and its essentially solar isotopic heritage to be preserved (Fig. 4). Of course, mass-selective argon loss from Mars via other mechanisms (e.g., solar wind sputtering) occurs and is ongoing (JAKOSKY et al., 2017). The hybrid mixture model provides a framework for assessing the relative importance of various mechanisms of atmospheric loss over the entire history of the planet.

## 3. Discussion

***The origin of the Martian hydrosphere.*** A significant feature of the Martian volatile record is that the surface hydrosphere inferred by geomorphology – like the atmospheric $^{36}$Ar reservoir – cannot be generated via volcanic outgassing over time (JAKOSKY AND TREIMAN, 2023). The hydrosphere was apparently placed on the Martian surface early in planetary history. Independent evidence for the existence of a Martian surface hydrosphere in the first 100 Myrs comes from an excess of $^{129}$Xe in the atmosphere from the decay of water-soluble and short-lived $^{129}$I (MUSSELWHITE *et al.*, 1991). The hybrid origin model for the Martian atmosphere suggests a new mechanism for the formation of a hydrosphere. Although we have considered the nebular and chondritic gases to be chemically inert, in reality, the H$_2$-dominated nebular gas can react with outgassed oxides (*e.g.*, CO$_2$) to produce new planetary water (H$_2$+CO$_2$→H$_2$O+CO). If the CO thus produced escapes as CO, there is a net gain of water at the Martian surface. Both the strong D/H enrichment of the early Martian hydrosphere (GREENWOOD *et al.*, 2008) and the anomalous oxygen recorded in ~4.43 billion year old zircons (NEMCHIN *et al.*, 2014) may result from isotopic exchange between a hydrosphere and an escaping H$_2$-dominated atmosphere (PAHLEVAN *et al.*, 2022; ZAHNLE AND KASTING, 2023).

***Cometary contribution to the inner Solar System.*** The low C/$^{36}$Ar and N/$^{36}$Ar of the Martian atmosphere relative to chondrites (Fig. 3) has previously been attributed to a possible contribution from comets, which are expected to be Ar-rich (MARTY *et al.*, 2016). However, a cometary origin introduces some problems even as it solves others. Results from the Rosetta mission allowed the identification of cometary xenon as a likely source for terrestrial atmospheric xenon, in particular the long-hypothesized component called U-Xe that is apparent in the atmosphere of Earth but not Mars (MARTY *et al.*, 2017). Assuming comet 67P/Churyumov-Gerasimenko – which Rosetta sampled – is representative of the cometary reservoir, the question arises as to why Martian atmospheric xenon does not record any signature of a cometary component. The resolution to this dilemma may be the relative retention of volatiles during impacts onto Earth and Mars. Cometary impacts onto terrestrial planets are high-velocity events sufficiently energetic to vaporize both icy and silicate components, producing impact vapor plumes. The fate of impact plumes (retained or lost) depends on the ratio of impact to escape velocity, such that cometary vapor plumes on Earth tend to be gravitationally retained whereas those on Mars tend to disperse from the weaker gravity field present (ZAHNLE, 1993). More work is needed to better understand the role of various escape processes in sculpting the volatile inventory of the terrestrial planets.

**Acknowledgements**

K.P. acknowledges support from NASA's Solar System Workings Program (80NSSC21K1833). The authors acknowledge discussions with Sujoy Mukhopadhyay and comments on an early draft from Kevin Zahnle, Alessandro Morbidelli, and James Lyons, which helped to improve the manuscript.

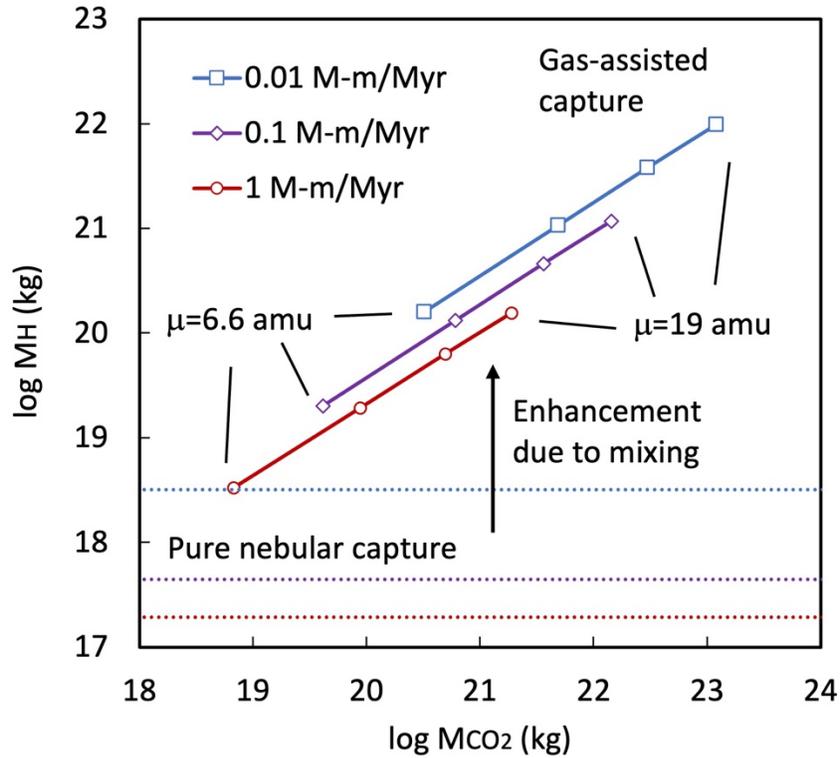

**Figure 1.** The captured gas inventory ($M_H$) as a function of the outgassed inventory ($M_{CO2}$) for fully-mixed hybrid atmospheres. Curves correspond to different accretion rates (in units of Mars masses per million years) and therefore to different planetary luminosities. "Pure nebular capture" (dotted lines) refer to nebular atmospheres with no outgassed component mixed in ($M_{CO2}=0$) for reference and is comparable to earlier calculations (ERKAEV *et al.*, 2014). Mixing with a high molecular weight component enhances nebular capture with the magnitude of the enhancement dependent on the mixed in heavy gas inventory. The closeness of the solid line slopes to unity indicates the near constant proportion of nebular to chondritic gases in the calculated atmospheres. The four data points along each curve correspond to a range of mean molecular weights characterizing fully-mixed hybrid atmospheres.

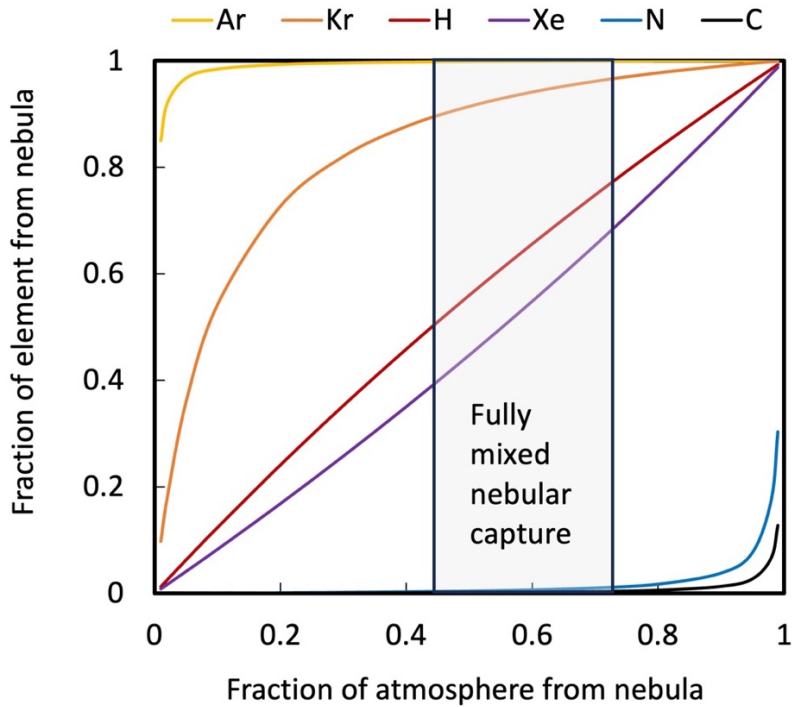

**Figure 2.** Mixtures of chondritic and solar-composition gases make specific predictions for the provenance of elements deriving from each component. The krypton isotopic composition of the Martian atmosphere – like solar and distinct from chondrites (PEPIN, 1991; CONRAD *et al.*, 2016) – suggests the nebular contribution to the primordial Martian atmosphere was >10% counting atoms. Upper limits on the contribution of the nebular component to Martian volatiles derive from the physics of nebular capture into fully mixed hybrid atmospheres and are <46-77% (see Fig. 1). Neon is not plotted for clarity but like Ar is derived almost entirely from the nebular component. Data used to make this plot is given in Table S-1.

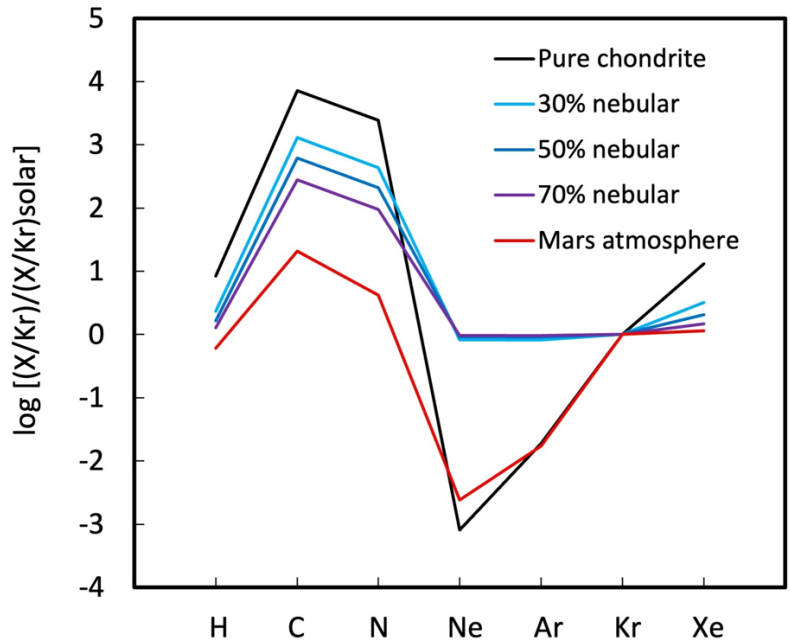

**Figure 3.** Relative abundances of volatile elements in mixtures of chondritic and solar composition gases normalized to krypton and to solar composition. Percentages refer to the fraction of total atoms in the mixture contributed by the solar component. Mixtures with varying proportions of the nebular component can be compared to the composition of the present-day Martian atmosphere and differences between the two compositions can be used to infer the imprint of loss processes. The similarity of the cyan, blue, and magenta curves attests to a well-constrained chemical composition for the initial hybrid atmosphere.

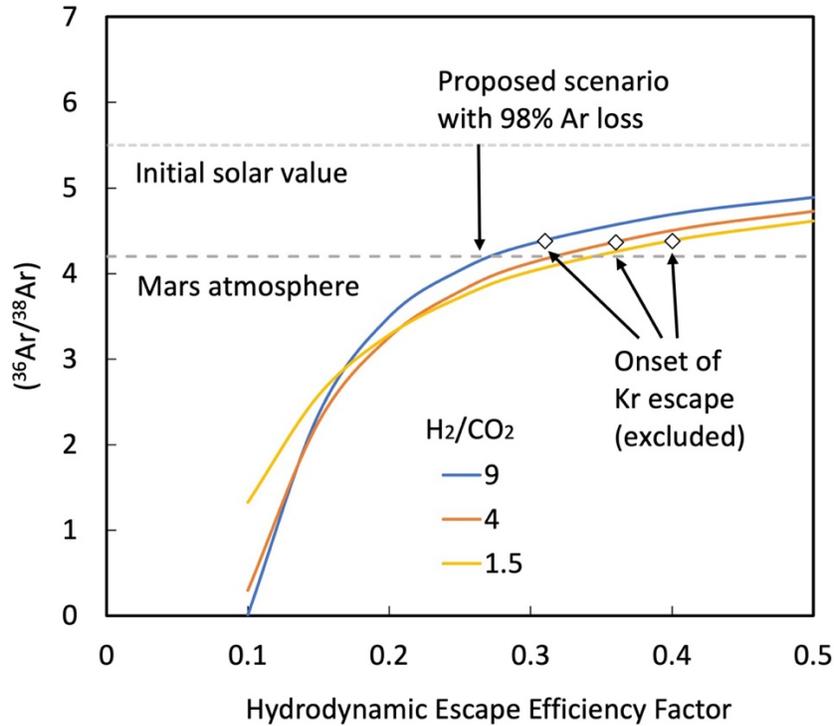

**Figure 4.** Mass-dependent fractionation accompanying argon loss from the hybrid Mars atmosphere. The solid lines show the argon isotopic composition of an atmosphere after Rayleigh distillation driven by hydrodynamic escape producing 50x Ar-depletion, for various nebular to chondritic proportions ($H_2/CO_2$), starting with solar Ar ($^{36}Ar/^{38}Ar=5.5$). The efficiency factor is a proxy for the mass-loss rate. Preservation of solar-like isotopic Kr requires Kr non-participation in the mass-fractionating outflow and sets upper limits on the efficiency factor. The existence of solutions sufficiently vigorous to allow large (50x) reductions in Ar/Kr (see Fig. 3) without excessive Ar (and no Kr) isotopic fractionation indicates the viability of a hybrid mixture as the initial composition of the Martian atmosphere.

# A hybrid origin for the Martian atmosphere

K. Pahlevan, L. Schaefer, D. Porcelli

## Supplementary Information

The Supplementary Information includes:
- Enhancement of Nebular Capture
- Energetics, Mechanisms, and Extent of Mixing
- Fractionation via Hydrodynamic Escape
- Magnitude of Impact Erosion
- Supplementary Tables S-1 and S-2
- Supplementary Information References

## Enhancement of Nebular Capture

**Pure nebular captured atmospheres**
The structure of a planetary atmosphere in hydrostatic equilibrium and thermal steady state with a gas disk is related to the planetary luminosity and the temperature ($T_H$) and density ($\rho_H$) of the gas disk in which it is embedded (HAYASHI *et al.*, 1979). For nebular conditions at the radial distance of Mars, we adopt $T_H = 200\ K$ and $\rho_H = 5 \cdot 10^{-10} g\ cm^{-3}$ (HAYASHI, 1981). We take the outer boundary of the atmosphere where the hydrostatic structure continuously connects to the solar nebula as the Hill radius, $R_H [\equiv a \cdot (M_p/3M_*)^{1/3}]$, with $a$ the semi-major axis of the planetary orbit, $M_p$ the planetary mass, and $M_*$ the stellar mass. $R_H$ for Mars is ≈320 planetary radii. We note that different model assumptions are possible (e.g., outer boundary at $R_H/4$) (CHAMBERS, 2017) but that choice of outer boundary condition has only a minor effect on calculated densities lower in the atmosphere where most of the mass resides (PISO AND YOUDIN, 2014).

The mass of the nebular-captured atmosphere is calculated using the minimum of the Hill radius and the Bondi radius $R_B [\equiv GM_p \bar{\mu}/k_B T_H]$ (IKOMA AND GENDA, 2006) with $G$ is the universal gravitational constant, $\bar{\mu}$ the mean molecular weight of the atmosphere (=2.4 amu for a $H_2$-He-rich nebular gas), and $k_B$ Boltzmann's constant. For nebular capture onto Mars, the Bondi radius is smaller than the Hill radius by about an order of magnitude. Hence, even though the Hill sphere is the outer boundary where hydrostatic equilibrium with the solar nebula is assumed, captured atmospheric masses are calculated by considering only the mass inside the Bondi radius.

To calculate vertical temperature and density structure, we consider compositionally uniform two-layer atmospheres in which the lower layer is convective and the upper layer is radiative  At the bottom of the atmosphere, we consider heat input via gravitational energy release due to planetesimal accretion:

$$L_{in} = \frac{GM_p \dot{M}_p}{R_p} \quad \text{(Eq. S-1)}$$

with $M_p$ and $R_p$ the planetary mass and radius, and $\dot{M}_p$ the mass accretion rate, which we take to be 0.01-1 Mars mass/Myr. $^{26}$Al heating may also be relevant for early Mars but can be modelled using this range of accretion rates. In quasi-steady state, this heat input is balanced by radiative heat losses at the radiative-convective boundary of the surrounding atmosphere (GINZBURG *et al.*, 2016):

$$L_{out} = \frac{64\pi\sigma_{SB}T_{rcb}^4 R'_B}{3\kappa_{rcb}\rho_{rcb}} \quad \text{(Eq. S-2)}$$

with $\sigma_{SB}$ the Stefan-Boltzmann constant, $T_{rcb}$ the temperature at the radiative-convective boundary, which is the same as the nebular temperature to within a factor of order unity (PISO AND YOUDIN, 2014), $\kappa_{rcb}$ and $\rho_{rcb}$ are the opacity and density at the radiative-convective boundary, and $R'_B$ is the modified Bondi radius:

$$R'_B = \frac{\gamma-1}{\gamma}\frac{GM_p\bar{\mu}}{k_B T_H} \quad \text{(Eq. S-3)}$$

with $\gamma$ the adiabatic index, for which we adopt a value appropriate for diatomic molecules (7/5) (KITTEL AND KROEMER, 1998). For our analysis, we make the approximation that $T_{rcb} = T_H$. To solve for the atmospheric structure, a prescription for the upper atmospheric opacity near the radiative-convective boundary is needed. We assume atmospheric opacity is dominated by dust with a size-distribution like that of the interstellar medium (BELL AND LIN, 1994):

$$\kappa = 2\cdot\left(\frac{T}{100\,K}\right)^2 \, cm^2 g^{-1} \quad \text{(Eq. S-4)}$$

With these equations, it is possible to solve for the density at the radiative-convective boundary ($\rho_{rcb}$) as a function of the planetesimal accretion rate ($\dot{M}_p$), which is one of the parameters that determines the total mass of captured gas. For conditions typical of the radiative-convective boundary ($\rho_{rcb}\sim 10^{-8} g\,cm^{-3}$, $T_{rcb}\sim 200\,K$, $P_{rcb}\sim 0.2\,millibars$), the opacity is dominated by the dust, with molecular lines and collision-induced absorption of $H_2$ contributing several orders of magnitude lower opacity, as evident in e.g., Figure 1 of (LEE *et al.*, 2014). Although opacity in deeper layers can be dominated by pressure-broadened gas absorption, the entropy of the adiabat is determined by conditions in the upper atmosphere where dust opacity dominates. In addition, we solve for the radial distance of the radiative-convective boundary ($R_{rcb}$). Approximating the radiative layer above the radiative-convective boundary as an isothermal layer in hydrostatic equilibrium to the Hill sphere, we obtain:

$$\frac{R_{rcb}}{R_B} = \left[\ln\left(\frac{\rho_{rcb}}{\rho_H}\right) + \frac{R_B}{R_H}\right]^{-1} \quad \text{(Eq. S-5)}$$

With these parameters, we solve for the radius of the radiative-convective boundary ($R_{rcb}$) as a function of the density at that location ($\rho_{rcb}$). Because atmospheric mass in the radiative layer is negligible due to the exponentially decreasing density in that region (PISO AND YOUDIN, 2014), the total mass of the nebular captured atmosphere can be expressed by integrating the density structure of the lower adiabatic convective layer (GINZBURG *et al.*, 2016):

$$M_H = \left(\frac{5\pi^2}{4}\right) R_{rcb}^3 \rho_{rcb} \left(\frac{R'_B}{R_{rcb}}\right)^{\frac{1}{\gamma-1}} \quad \text{(Eq. S-6)}$$

**Fully-mixed hybrid atmospheres**

A fully-mixed hybrid atmosphere is one in which the outgassed components (e.g., $CO_2$) and the nebular captured gases are present in uniform proportions throughout the Hill sphere. A fully-mixed state describes the maximum enhancement of the nebular captured gaseous inventory and is therefore a useful and easily calculable end-member state. Identical expressions for the incoming luminosity ($L_{in}$), outgoing luminosity ($L_{out}$), and modified Bondi sphere ($R'_B$) (Eq. S-1, S-2, and S-3) can be used. An important difference is that the mean molecular weight within the Hill sphere is no longer equal to that of the surrounding nebular environment but is given by the proportion of outgassed and nebular captured components:

$$\bar{\mu} = x_H\mu_H + (1-x_H)\mu_{CO2} \quad \text{(Eq. S-7)}$$

where $x_H$ represents the mole fraction of gas species that derive from the nebular component, with $\mu_H$ and $\mu_{CO2}$ equal 2.4 and 44 amu, respectively. Two other adjustments to the capture model must be made to accommodate uniform mixing with an outgassed heavy gas component. First, the opacity law we have adopted (Eq. S-4) is appropriate to dust-dominated opacity, which may be the case for the nebular captured component but not the outgassed component, which is thought to be exsolved from a magma ocean and is assumed to be dust-free. Accordingly, we assume that the opacity of the mixture is dominated by the opacity of dust in the nebular component, and adopt an expression for a lower opacity relative to a pure nebular atmosphere:

$$\kappa = 2 \cdot \left(\frac{T}{100\ K}\right)^2 \cdot x_H \cdot \left(\frac{\mu_H}{\bar{\mu}}\right)\ cm^2 g^{-1} \qquad \text{(Eq. S-8)}$$

The procedure for calculating fully-mixed hybrid atmosphere solutions is to take the mole fraction of the nebular component ($x_H$) as an independent variable and select a value, which in practice is in the range 0.5-1. With the composition of the hybrid mixture specified, Equations S-1 to S-3, S-7, and S-8 can be used to solve for the density at the radiative-convective boundary ($\rho_{rcb}$). As with the pure nebular case, the mass of the atmosphere is dominated by the mass in the convective layer. Accordingly, the radius of the radiative-convective boundary must be specified to calculate captured gas masses. A fully-mixed hybrid atmosphere is distinct from the pure nebular atmosphere because it hosts a compositional boundary at the Hill radius between the hybrid mixture and the nebular-composition gas. Hence, a relationship between the conditions across the compositional boundary must be specified. Hydrostatic equilibrium and thermal steady-state requires that pressure and temperature be continuous across the compositional boundary. For ideal gases, these requirements translate to a relationship between the density of the hybrid atmosphere approaching the Hill sphere ($\rho_{out}$) and that of the external nebula ($\rho_H$):

$$\rho_{out} = \rho_H \cdot \frac{\mu_H}{\bar{\mu}} \qquad \text{(Eq. S-9)}$$

The radius of the radiative-convective boundary ($R_{rcb}$) can then be expressed by an equation analogous to Eq. S-5:

$$\frac{R_{rcb}}{R_B} = \left[\ln\left(\frac{\rho_{rcb}}{\rho_{out}}\right) + \frac{R_B}{R_H}\right]^{-1} \qquad \text{(Eq. S-10)}$$

Because the mass in the low density radiative layer is expected to be negligible (PISO AND YOUDIN, 2014), the mass of both the nebular captured gas and the outgassed inventory can be expressed by integrating the density structure across the dense adiabatic convective layer (GINZBURG et al., 2016) and multiplying by the mass fraction deriving from the nebular and outgassed components, respectively:

$$M_H = x_H \cdot \left(\frac{\mu_H}{\bar{\mu}}\right) \cdot \left(\frac{5\pi^2}{4}\right) R_{rcb}^3 \rho_{rcb} \left(\frac{R_B'}{R_{rcb}}\right)^{\frac{1}{\gamma-1}} \qquad \text{(Eq. S-11)}$$

$$M_{CO2} = (1 - x_H) \cdot \left(\frac{\mu_{CO2}}{\bar{\mu}}\right) \cdot \left(\frac{5\pi^2}{4}\right) R_{rcb}^3 \rho_{rcb} \left(\frac{R_B'}{R_{rcb}}\right)^{\frac{1}{\gamma-1}} \qquad \text{(Eq. S-12)}$$

## Energetics, Mechanisms, and Extent of Mixing

Mixing a stably-stratified fluid with a high-density deep outgassed layer and low-density shallow nebular component – as envisioned in e.g., (SAITO AND KURAMOTO, 2018) – into a fully-mixed structure considered in this work requires an energy source and a fluid dynamical mechanism to channel that energy into mixing. We briefly discuss these considerations in turn. Towards the end of Martian accretion ($M_p \approx M_{Mars}$) the compositional boundary in a stratified atmosphere between an outgassed layer and a nebular layer is ≈1-2 planetary radii, depending on the outgassed volatile inventory. By contrast, in the fully-mixed hybrid atmospheres we calculate in Figure 1, the radius of the radiative-convective boundary ($R_{rcb}$), where most of the atmospheric mass resides, is ≈8.4-46.5 planetary radii. Mixing a heavy outgassed component therefore requires nearly as much energy as complete removal of this component from the planetary potential well. To within a factor of about two, the energy required to uniformly mix an outgassed heavy component into a hybrid atmosphere is:

$$W_{grav} \cong \frac{GM_p M_{CO2}}{R_p} \qquad \text{(Eq. S-13)}$$

with $W_{grav}$ the work done in lifting the heavy gas into the fully-mixed state. Because the mass of the outgassed inventory is always much less than the planetary mass for terrestrial planets, and because this mixing occurs in the context of accretionary energy budgets of order $GM_p^2/R_p$, it is clear that the energy for mixing was present during accretion. A uniform composition hybrid atmosphere requires a fluid dynamical mechanism to channel ≈10⁻⁴ of the accretional energy into mixing. Although dust-free layered atmospheric models sometimes place the compositional boundary above the convective region and predict little convective mixing (SAITO AND KURAMOTO, 2018), dusty nebular atmospheres are more opaque and more likely to host an initial compositional boundary within the convective region. Because all gases are miscible, this picture of convective mixing is analogous to the problem of core erosion in the giant planets, where more dramatic heavy element redistribution scenarios have been considered (STEVENSON, 1982). How much convective mixing actually takes place is difficult to predict from first principles. We suggest that the cosmochemical record (Fig. 2) can be used to empirically determine the degree of mixing on Mars.

A constraint on the extent of mixing in the accretionary Mars system is the apparent preservation of chondritic mantle volatiles in the presence of nebular contributions to the atmosphere. The persistence of the mantle-atmosphere volatile dichotomy may indicate one of several possible histories. First, it is possible that significant portions of silicate Mars may have already solidified at the time of the nebular capture, effectively isolating chondritic mantle volatiles from mixing with the captured atmosphere. Secondly, although we have considered fully-mixed hybrid atmospheres for the calculations of nebular capture, it is possible that the compositional gradient in the accretionary atmosphere was smooth, with a $CO_2$-dominated lower atmosphere gradually transitioning to an $H_2$-dominated upper atmosphere, effectively sealing off the silicate planet from nebular additions. Finally, it is possible that there once was complete homogenization of the silicate-atmosphere hybrid volatile system on Mars, and that the chondritic character observed in the Martian mantle is the overprint of chondritic volatiles delivered during late accretion after the solidification and effective isolation of silicate Mars. At present, we have no way to clarify these scenarios other than to state that the persistence of the mantle-atmosphere dichotomy places constraints on the sequence of volatile acquisition, mixing, and isolation in the accretionary Mars system.

## Fractionation via Hydrodynamic Escape

The current $^{36}Ar/^{84}Kr$ ratio in the Martian atmosphere is lower than that of a hybrid mixture by a factor of ≈50 (Fig. 3). To illustrate the capacity of extreme-ultraviolet (EUV) powered hydrodynamic escape of a primordial volatile inventory to generate large depletions in argon with only modest isotopic fractionation, we consider a simple model in which multiple major species are present (ZAHNLE et al., 1990; ZAHNLE AND KASTING, 2023). We arbitrarily define "major" species as those with mole fractions greater than 10%, and adopt a two-component ($H_2$-$CO_2$) model, neglecting a possible role for CO, $CH_4$, $N_2$, and He, which we assume would be present in minor abundances. After the crystallization of the magma ocean, $H_2O$ would condense in the lower atmosphere (PAHLEVAN et al., 2022) such that water vapor is not expected to play a role in the loss processes of interest. Two-component hydrodynamic escape can be described with an equation of energy balance:

$$\phi_1 m_1 + \phi_2 m_2 = \frac{1}{4} \eta_{eff} S_{euv} \frac{R_p}{GM_p} \qquad \text{(Eq. S-14)}$$

with $\phi_i$ [cm$^{-2}$ s$^{-1}$] the number flux of component i out of the atmosphere, $m_1 (= 2\ amu)$ and $m_2 (= 44\ amu)$ the atomic mass of the two major constituents $\eta_{eff}$ the efficiency factor, which is a parameter (0-1) that represents the fraction of incoming EUV energy that is channelled into mass-loss, $S_{euv}$ [erg cm$^{-2}$ s$^{-1}$] is the EUV flux of the young Sun at the top of the atmosphere, and the factor of 4 reflects a spherical average. $\eta_{eff}$ embodies all uncertainties of the energy budget into one parameter. According to (ZAHNLE AND KASTING, 2023), $S_{euv}$ is 133 erg cm$^{-2}$ s$^{-1}$ at Mars's heliocentric distance for the first few tens of millions of years of Solar System history. For these calculations, we adopt this value as a constant.

Solving for the two fluxes requires one additional constraint. Above the homopause, molecular diffusion exceeds eddy diffusion – by definition – such that the different species can partially separate via diffusion and assume different scale heights. In hydrostatic atmospheres, this separation populates the upper atmosphere in lighter species allowing for mass-fractionation accompanying non-thermal escape processes (JAKOSKY et al., 2017). Such diffusive separation by mass also occurs in hydrodynamic outflows. One way to make the problem tractable is to assume an isothermal outflow, for which one can write the approximate relation (ZAHNLE AND KASTING, 2023):

$$\phi_1 \left(1 + \frac{x_2}{x_1}\right) - \phi_2 \left(1 + \frac{x_1}{x_2}\right) = \frac{g(m_2 - m_1)b_{12}}{k_B T} \qquad \text{(Eq. S-15)}$$

with $x_1$ and $x_2$ the mole fraction of the 1$^{st}$ and 2$^{nd}$ component, $g$ the gravitational acceleration, $b_{12}$ [cm$^{-1}$ s$^{-1}$] the binary diffusion coefficient of the gas pair, $k_B$ Boltzmann's constant, and $T$ the temperature of the escaping region. For concreteness, we adopt $T = 1,000\ K$ for all escape calculations. Binary diffusion coefficients measure the ease with which one gas diffuses through another. The values for binary diffusion coefficients for all gas pairs used in this work are listed in Table S-2. With Eq. S-14 and S-15, we can solve for the flux of H$_2$ ($\phi_1$) and CO$_2$ ($\phi_2$) as a function of the hydrodynamic escape efficiency parameter $\eta_{eff}$.

With fluxes of major species specified, the passive response of noble gases as witnesses of events can be described. For an increasingly vigorous outflow, increasingly massive gaseous species can be accelerated to space via frequent collisions with the outgoing gases. The outgoing flux ($\phi_j$) of a trace species with given atomic mass ($m_j$) and mole fraction ($x_j$) can, in the isothermal approximation, be written (ZAHNLE AND KASTING, 2023):

$$\frac{\phi_j}{x_j}\left(\frac{x_1}{b_{1j}} + \frac{x_2}{b_{2j}}\right) = \frac{g(m_2 - m_j)}{k_B T} + \frac{\phi_2}{x_2}\left(\frac{x_1}{b_{12}} + \frac{x_2}{b_{2j}}\right) + \frac{\phi_1}{x_1}\left(\frac{x_1}{b_{1j}} - \frac{x_2}{b_{12}}\right) \quad \text{(Eq. S-16)}$$

To calculate isotopic fractionation, an equation like Eq. S-16 is written for each of the isotopic species, keeping in mind that in some cases the binary diffusion coefficient of two gases noticeably changes upon isotopic substitution (Table S-2). The fractionation factor ($\alpha$) is the ratio of the ($^{36}$Ar/$^{38}$Ar) of the outflowing gas relative to that of the atmosphere from which it is sourced:

$$\alpha = \frac{(\phi_{36Ar}/x_{36Ar})}{(\phi_{38Ar}/x_{38Ar})} \quad \text{(Eq. S-17)}$$

With an expression for the fractionation factor ($\alpha$), the evolution of the Martian atmosphere $^{36}$Ar/$^{38}$Ar ratio (R$_{final}$) as a function of the initial ratio (R$_{initial}$) and the fraction of $^{36}$Ar remaining ($F$) at the conclusion of the hydrodynamic episode can be related with the Rayleigh fractionation formula:

$$\text{R}_{final} = \text{R}_{initial} F^{\alpha - 1} \quad \text{(Eq. S-18)}$$

In the case of Kr, no mass-dependent fractionation is detectable in the Martian atmosphere (PEPIN, 1991). The hybrid mixture model requires that Mars preserve its nebular heritage by experiencing negligible mass-selective Kr escape for the duration of the hydrodynamic episode. In the context of the adopted two-component (H$_2$-CO$_2$) model, this leads to the requirement that the escape fluxes ($\phi_1, \phi_2$) remain below their critical values for Kr entrainment. The critical fluxes can be expressed by setting $\phi_j$=0 in Eq. S-16. The corresponding joint constraint on $\phi_1$ and $\phi_2$ can be used, in concert with Eq. S-14 and S-15, to define the conditions for the onset of Kr escape. Importantly, there is a range of outgoing fluxes sufficiently low as to retain Kr while sufficiently high to lose $^{36}$Ar and $^{38}$Ar indiscriminately enough to be consistent with the observations (Fig. 4). We propose these fluxes characterize losses from the hybrid atmosphere.

Finally, because the diffusion properties of Kr and Xe are similar, Kr retention also implies Xe retention. Hence, the Xe loss required to reproduce the Kr/Xe ratio (Fig. 3) and Xe isotopic mass-fractionation in the Martian atmosphere must be due to another process. By measurement of trapped atmospheric gases in Martian meteorites, it has recently been inferred that the isotopic evolution of Martian Xe took place over hundreds of millions of years (CASSATA *et al.*, 2022). As on the Earth, where Xe isotopic evolution apparently persisted for two gigayears (AVICE *et al.*, 2018), the Xe loss process may have involved ionization (ZAHNLE *et al.*, 2019). If so, the xenon loss may have been decoupled from the hydrodynamic loss episode of the hybrid Martian atmosphere as neutral atoms and molecules calculated here.

## Magnitude of Impact Erosion

A solar-like krypton isotopic composition of the Martian atmosphere argues against substantial mass-selective loss of krypton. Given that krypton escape from present-day Martian is thought to be negligible (KUROKAWA *et al.*, 2021), and that the hydrodynamic escape episode we have proposed for massive argon loss would have excluded krypton, the question arises as to what process could have depleted the krypton inventory from its massive initial endowment down to its observed abundance in the rarefied Martian atmosphere. The only known loss process capable of eliminating atmospheric gases indiscriminately with respect to mass is impact erosion (MELOSH AND VICKERY, 1989; ZAHNLE, 1993; BRAIN AND JAKOSKY, 1998; SHORTTLE *et al.*, 2024), a type of hydrodynamic outflow that occurs on dynamical timescales. If the acquisition of solar-like krypton was due to Martian capture via mixing in a solar nebula component, as we propose, then the initial krypton inventory can be used to place constraints on the magnitude of impact erosion in sculpting the Martian atmospheric inventory over its entire history. As an example, we consider a nebular capture scenario in which Mars acquires $1.2 \times 10^{20}$ kg of solar-composition gas, equivalent to 30 bars of a $H_2$-He-rich atmosphere at the Martian surface (Fig. 1). A solar composition gas is composed of 80 parts per billion by weight (ppbw) krypton (LODDERS, 2003), corresponding to a primordial krypton inventory of $9.6 \times 10^{12}$ kg, to be compared with the present-day krypton inventory of $1.5 \times 10^{10}$ kg. This comparison suggests that non-fractionating impact erosion must have drawn down the Martian atmospheric volatile inventory by nearly three orders of magnitude, a conclusion regarding early losses previously reached using independent arguments (ZAHNLE, 1993; MARTY AND MARTI, 2002).

## Supplementary Tables

Table S-1  Adopted elemental abundances in relevant cosmochemical reservoirs

| Element | H chondrite[1,2] | EH chondrite[1,2] | Chondritic mixture[3] (55% H 45% EH) | Solar[4] | Martian atmosphere[5,6] |
|---|---|---|---|---|---|
| H | 0.000459 | 0.001309 | 0.0008415 | 1E12 | 16.2 |
| C | 0.0001 | 0.000342 | 0.0002088 | 2.88E8 | 0.16 |
| N | 2.43E-6 | 0.00004 | 1.934E-05 | 7.94E7 | 8.8E-3 |
| Ne | 1.00E-11 | 3.80E-12 | 7.210E-12 | 8.91E7 | 6.3E-7 |
| Ar | 1.11E-12 | 1.63E-11 | 7.961E-12 | 4.17E6 | 1.9E-6 |
| Kr | 2.14E-13 | 2.50E-13 | 2.304E-13 | 2.29E3 | 6.1E-8 |
| Xe | 3.82E-14 | 6.11E-13 | 2.958E-13 | 2.24E2 | 6.8E-9 |

[1] H, C, N (SCHAEFER AND FEGLEY, 2017)
[2] Ne, Ar, Kr, Xe (SCHULTZ et al., 1991; PATZER AND SCHULTZ, 2002)
[3] Chondritic mixture model for Mars (SANLOUP et al., 1999)
[4] Corrected for heavy element settling in the Sun (LODDERS, 2003)
[5] H from a ≈500 m global equivalent layer (GEL) of water (DI ACHILLE AND HYNEK, 2010)
[6] C, N, Ne, Ar, Kr, Xe from (HALLIDAY, 2013)

Table S-2  Binary diffusion coefficients for hydrodynamic escape model

| Gas pair | b (cm$^{-1}$ s$^{-1}$) | Source or scaling |
|---|---|---|
| $H_2$-$CO_2$ | $4.1 \times 10^{19}$ $(T/1000)^{0.75}$ | Zahnle and Kasting (1986) |
| $^{36}$Ar-$H_2$ | $5.0 \times 10^{19}$ $(T/1000)^{0.75}$ | Zahnle and Kasting (1986) |
| $^{38}$Ar-$H_2$ | $5.0 \times 10^{19}$ $(T/1000)^{0.75}$ | Scaled from $^{36}$Ar-$H_2$ |
| $^{84}$Kr-$H_2$ | $4.4 \times 10^{19}$ $(T/1000)^{0.76}$ | Zahnle and Kasting (1986) |
| $^{36}$Ar-$CO_2$ | $1 \times 10^{19}(T/1000)^{0.75}$ | Zahnle and Kasting (2023) |
| $^{38}$Ar-$CO_2$ | $0.985 \times 10^{19}(T/1000)^{0.75}$ | Scaled from $^{36}$Ar-$CO_2$ |
| $^{84}$Kr-$CO_2$ | $0.78 \times 10^{19}(T/1000)^{0.75}$ | Scaled from $^{36}$Ar-$CO_2$[1] |

[1] Scaling across elements requires kinetic diameters.: Ar: 340 pm; Kr: 360 pm; $CO_2$: 330 pm

## Supplementary Information References